\documentclass[11pt,a4paper]{article}

\usepackage[a4paper,margin=25mm]{geometry}
\usepackage{graphicx}
\usepackage{amsmath,amssymb}
\usepackage{authblk}
\usepackage{caption}
\usepackage{subcaption}
\usepackage{hyperref}
\usepackage[numbers,sort&compress]{natbib}

\graphicspath{{figures/}}

\title{Constraint-Enhanced Physical Search through Correlation Matching}

\author[1]{Song-Ju Kim\thanks{Email: kim@sobin.org}}
\affil[1]{SOBIN Institute LLC, Kawanishi, Hyogo, Japan}

\begin{document}

\maketitle

\begin{abstract}
  Physical systems do not merely add noise to search processes; they impose constraints that generate structured correlations.  We propose a principle of constraint-enhanced physical search in which temporal correlations in exploration are matched to constraint-induced spatial correlations in the update dynamics.  Using a minimal tug-of-war bandit model (TOW), we show that a conservation law converts local observations into differential evidence across alternatives, while a temporally correlated drive controls the order of exploration.  Search efficiency is improved not by stronger randomness or by maximal anti-correlation, but by matching the temporal correlation to the physical update scale that converts feedback into evidence.  A scaling estimate identifies the update-noise-to-contrast ratio as the leading parameter that limits how strongly temporal anti-correlation can be used.  The results suggest a general organizing principle for physical search: constraints and fluctuations can generate structured spatiotemporal correlations, and efficient exploration emerges when these correlations are matched to the update dynamics.
\end{abstract}

\section{Introduction}

Many difficult computational problems require efficient search, although the origin of the difficulty differs from problem to problem.  In combinatorial problems such as satisfiability problems (SAT) and route-search problems such as the traveling-salesman problem (TSP), the number of possible configurations grows rapidly, often exponentially, with system size \cite{GareyJohnson1979}.
In multi-armed bandit problems, by contrast, the difficulty is not primarily combinatorial explosion but the exploration--exploitation dilemma: one must gather information about uncertain alternatives while also exploiting the currently best option \cite{Robbins1952,Auer2002,LattimoreSzepesvari2020}.
Despite these differences, all of these problems require a search process that avoids redundant exploration and converts limited local information into useful guidance.

A central aim of physical and bio-inspired computing, often discussed within the broader context of natural computing, is to use the properties of physical systems to realize such efficient search.  Physical systems are not merely implementations of abstract algorithms.  They can update many degrees of freedom in parallel, evolve continuously in time, exploit fluctuations, and impose conservation laws or constraint-induced couplings among variables.  These features can reshape search dynamics by propagating local information, suppressing redundant trials, and coordinating the motion of alternatives or variables.  The question is therefore not only whether a physical system can represent a search problem, but what physical properties generate the efficiency of the search.

This viewpoint underlies previous physical approaches to difficult search and decision problems.
Tug-of-war (TOW) dynamics was introduced as a resource-conserving search principle for bandit problems \cite{KimTOW2010, Kim2015} and later implemented in physical decision-making systems \cite{Kim2013,Naruse2015,Naruse2017}.
Amoeba-inspired approaches to TSP search exploit correlated deformation and resource redistribution in a physical or biological substrate \cite{Zhu2013,AmoebaTSP}.  Amoeba-inspired SAT solvers and ratchet-based implementations further suggest that constrained physical dynamics can coordinate many local decisions in parallel \cite{AmoebaSAT,AmoebaSAT2}.
Chaotically driven and correlated search processes show that temporal structure can also reduce redundant exploration \cite{Kim2016,Naruse2017}; a simple temporal-only illustration is given in Appendix C.
These examples motivate a common question: {\it \bf what is the minimal physical principle by which structured dynamics improves search?}

The aim of the present work is to extract such a principle in the simplest setting where the relevant components can be separated.  We focus on two ingredients that repeatedly appear in physical search systems: temporally structured exploration and constraint-induced redistribution of local information.
Temporal structure controls the order in which alternatives are explored and can reduce redundant sampling.
Constraint-induced redistribution converts a local observation into a coordinated update across alternatives.  Our central claim is that efficient physical search emerges when these two correlation structures are compatible.

In this paper we study this principle using a minimal TOW bandit model.  We do not claim that the TOW dynamics represents the full complexity of SAT or TSP dynamics.  Rather, TOW provides a controllable reduced system in which temporal correlations and conservation-induced option-wise correlations can be varied independently and measured cleanly.  The temporal component is represented by a correlated exploratory drive.
The spatial component is represented by a conservation-like TOW update that converts reward feedback into an anti-correlated update of competing alternatives.
By combining these two components, we ask when temporal exploration is efficiently converted into differential evidence, and we derive a scaling estimate for the optimal temporal anti-correlation.

The resulting principle is not that more randomness is better, nor that stronger temporal anti-correlation is always better.  Useful temporal structure must be tuned to the physical update scale through which local feedback is converted into coordinated evidence.
In this sense, the present work addresses a physical source of efficiency in search: constraints and fluctuations generate structured correlations, and efficient search emerges from their interaction.

The contribution of this paper is threefold.  First, we formulate a minimal TOW bandit model in which the temporal exploration drive and the conservation-induced update can be varied independently.  Second, we introduce an interaction diagnostic that separates the independent effects of temporal anti-redundancy and TOW-induced differential evidence formation.  Third, we derive a scaling estimate showing that the constructive use of temporal anti-correlation is limited by the update-noise-to-contrast ratio.

\section{Minimal TOW bandit model}

Consider two alternatives \(A\) and \(B\) with reward probabilities \(P_A>P_B\).
The reward gap is
\begin{equation}
\Delta P=P_A-P_B .
\end{equation}
The internal values are \(Q_A,Q_B\), and the differential evidence is
\begin{equation}
Z=Q_A-Q_B .
\end{equation}
A temporally correlated binary signal \(s_t\in\{-1,+1\}\) modulates the choice boundary:
\begin{equation}
a_t=
\begin{cases}
A, & Z_t+\sigma s_t > 0,\\
B, & Z_t+\sigma s_t < 0.
\end{cases}
\label{eq:decision_rule}
\end{equation}
At the zero boundary, \(Z_t+\sigma s_t=0\), the choice is randomized.
The signal obeys a two-state Markov dynamics
\begin{equation}
\Pr(s_{t+1}=s_t)=\frac{1+\lambda}{2},
\qquad
\Pr(s_{t+1}=-s_t)=\frac{1-\lambda}{2}.
\label{eq:signal}
\end{equation}
Thus \(\lambda<0\) represents temporal anti-correlation.
In this minimal signal model we use
\begin{equation}
\chi=\max(0,-\lambda)
\end{equation}
as a proxy for temporal non-redundancy.
For a general physical signal, this can be replaced by a weighted correlation functional
\begin{equation}
\chi=-\sum_{\tau\ge1}w_\tau C_s(\tau),
\label{eq:chi_general}
\end{equation}
where \(C_s(\tau)\) is the task-relevant autocorrelation.

The TOW component is implemented as a generalized conservation-induced anti-correlated update.
If the selected option receives
\begin{equation}
d_t=
\begin{cases}
+\eta, & R_t=1,\\
-\eta\omega_0, & R_t=0,
\end{cases}
\end{equation}
then the competing option receives \(-\kappa d_t\).  For example, when \(A\) is selected,
\begin{equation}
Q_A\leftarrow Q_A+d_t,
\qquad
Q_B\leftarrow Q_B-\kappa d_t .
\label{eq:update_A}
\end{equation}
Here \(\kappa=0\) is an ordinary one-sided update, whereas \(\kappa=1\) gives exact TOW conservation,
\begin{equation}
Q_A+Q_B=\mathrm{const.}
\end{equation}
The penalty parameter is the TOW value
\begin{equation}
\omega_0=\frac{\gamma}{2-\gamma},
\qquad
\gamma=P_A+P_B,
\label{eq:omega0}
\end{equation}
unless stated otherwise.
The differential evidence update is amplified as
\begin{equation}
\Delta Z=(1+\kappa)d_t .
\label{eq:dZ}
\end{equation}
The signal \(s_t\) does not directly update \(Z_t\).
Instead, it modulates which alternative is sampled through the choice boundary in Eq.~\eqref{eq:decision_rule}.
During the early phase \(|Z_t|<\sigma\), this modulation changes the temporal balance of visits to the two alternatives.
The resulting reward observation determines \(d_t\), and the TOW rule then converts this local feedback into the differential update \(\Delta Z=(1+\kappa)d_t\).
Thus the temporal drive shapes the order of observations, while the conservation-induced update determines how each observation is accumulated as differential evidence.

\begin{figure}[!t]
\centering
\includegraphics[width=0.98\textwidth]{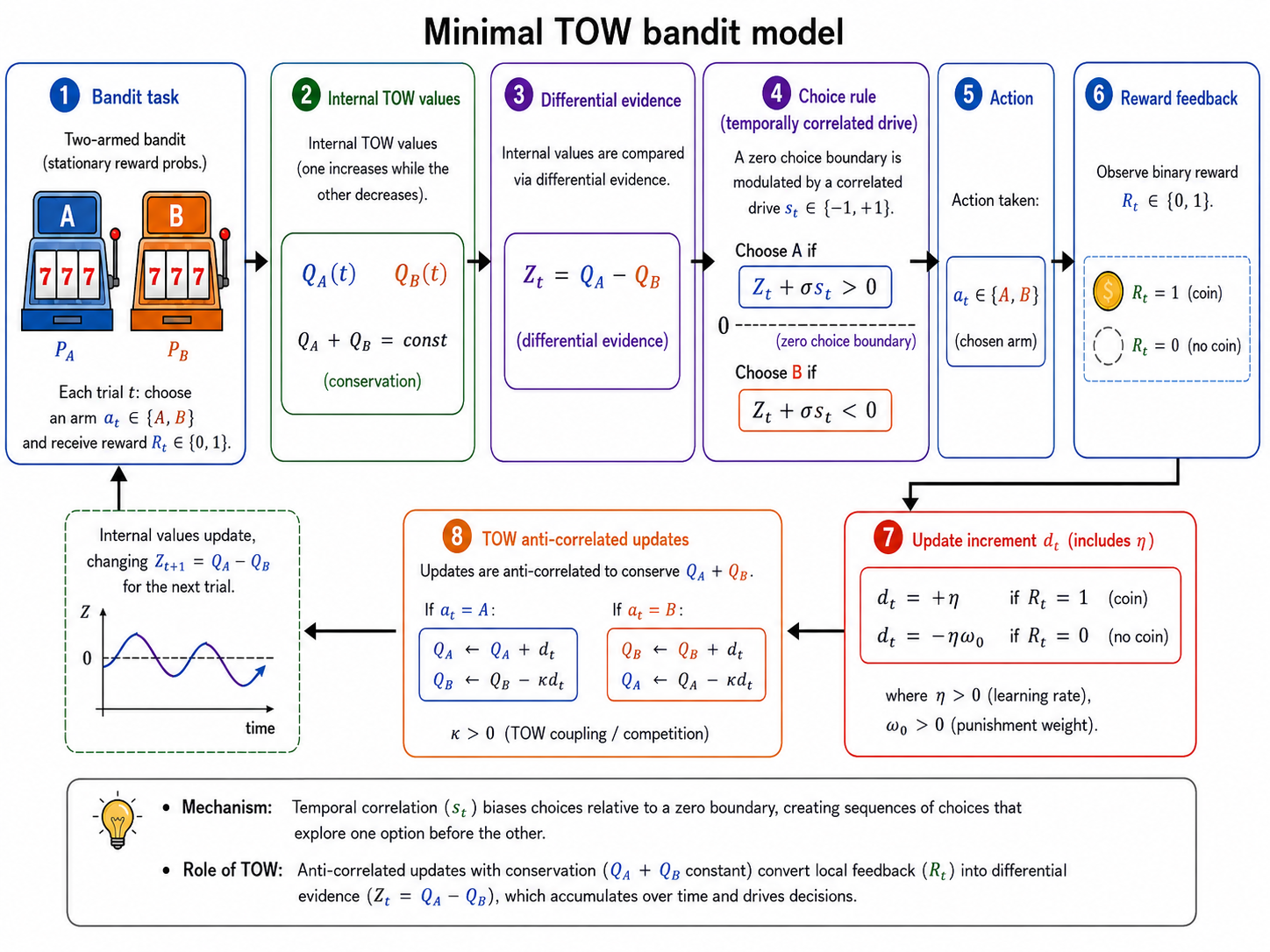}
\caption{
Minimal TOW bandit model.
A two-armed bandit trial consists of choosing an arm \(a_t\in\{A,B\}\), observing a binary reward \(R_t\in\{0,1\}\), and updating the internal TOW values \(Q_A\) and \(Q_B\).
For \(\kappa=1\), the update gives exact TOW conservation \(Q_A+Q_B=\mathrm{const}\).
The differential evidence \(Z_t=Q_A-Q_B\) is compared with a zero choice boundary after modulation by the temporally correlated drive \(\sigma s_t\).
Reward feedback defines the update increment \(d_t\), and the TOW rule updates the chosen arm and the competing arm in opposite directions.
Thus a local reward observation is converted into an option-wise anti-correlated update of the internal state.
}
\label{fig:tow_concept}
\end{figure}

Figure~\ref{fig:tow_concept} summarizes the minimal TOW bandit model.  A two-armed bandit trial consists of choosing an arm, observing a binary reward, and updating the internal TOW values.  The figure also shows how the temporally correlated drive shifts the zero choice boundary and how reward feedback is converted into an anti-correlated update of the internal values.
The variables \(Q_A\) and \(Q_B\) should be understood as internal physical evidence variables, not as direct estimates of \(P_A\) and \(P_B\).  The decision is made from their difference \(Z_t=Q_A-Q_B\), while the exploratory drive \(\sigma s_t\) shifts the effective choice boundary.  The TOW rule then converts a local reward or no-reward event into an anti-correlated update of the two internal values.

The same choice--update cycle can also be represented as a discretized finite-state Markov process, as described in Appendix~\ref{app:finite_state}.  This formulation is not used to replace the scaling argument, but it shows that the minimal model has a well-defined stochastic dynamics from which regret can be computed directly after discretization.

\paragraph{Interaction between temporal and spatial correlations.}
Having defined the choice and update cycle, we next quantify whether the temporal drive and the TOW update act independently or constructively.
Here the spatial component refers to the option-wise anti-correlation generated by the TOW constraint, or equivalently to the constraint-induced redistribution of local feedback across alternatives.
Let \(R_{\chi\kappa}\) be the regret for a system with temporal anti-correlation \(\chi\) and option-wise anti-correlation \(\kappa\).

With \(R_{00}\) as the uncorrelated baseline, define
\begin{equation}
{\cal E}_{\chi\kappa}=\frac{R_{00}}{R_{\chi\kappa}}.
\end{equation}
To isolate the interaction between temporal and spatial correlations, define
\begin{equation}
I(\chi,\kappa)=
{\cal E}_{\chi\kappa}
-{\cal E}_{\chi0}
-{\cal E}_{0\kappa}
+1,
\label{eq:interaction}
\end{equation}
and
\begin{equation}
\Lambda(\chi,\kappa)=\frac{I(\chi,\kappa)}{\chi\kappa}.
\label{eq:Lambda}
\end{equation}
Positive \(\Lambda\) means that temporal anti-correlation is not merely adding independent exploration; it is being converted by the TOW conservation law into useful differential evidence.

\section{Constructive synthesis and intermediate anti-correlation}

The simulations show a robust efficiency-enhancing regime in the combined system, together with a synthesis diagnostic that quantifies the non-additive part of the temporal--spatial interaction.
The effect is not obtained by making the signal maximally anti-correlated.
For the representative setting
\[
P_A=0.7,
\quad
P_B=0.5,
\quad
\sigma\simeq1.3,
\quad
\eta=0.5,
\quad
\kappa=1,
\]
the parameters lie in a representative transduction regime: the exploratory boundary shift is large enough to affect the early choice sequence, while the update step is not so large that the dynamics immediately locks into an early noisy decision.  In this regime, the temporal drive, reward feedback, and TOW update all contribute measurably, allowing the interaction between temporal anti-redundancy and constraint-induced differential evidence formation to be diagnosed.

The data show a broad efficiency-enhancing region at intermediate negative correlations.  We therefore focus on the robust improvement of the combined system and on the synthesis diagnostic, rather than on the precise location of local extrema in \(\lambda\).  This distinguishes correlation matching from a simple anti-correlation principle: complete alternation is not optimal because the temporal drive must match the spatial update scale.

\begin{figure}[!t]
  \centering
  \includegraphics[width=0.98\linewidth]{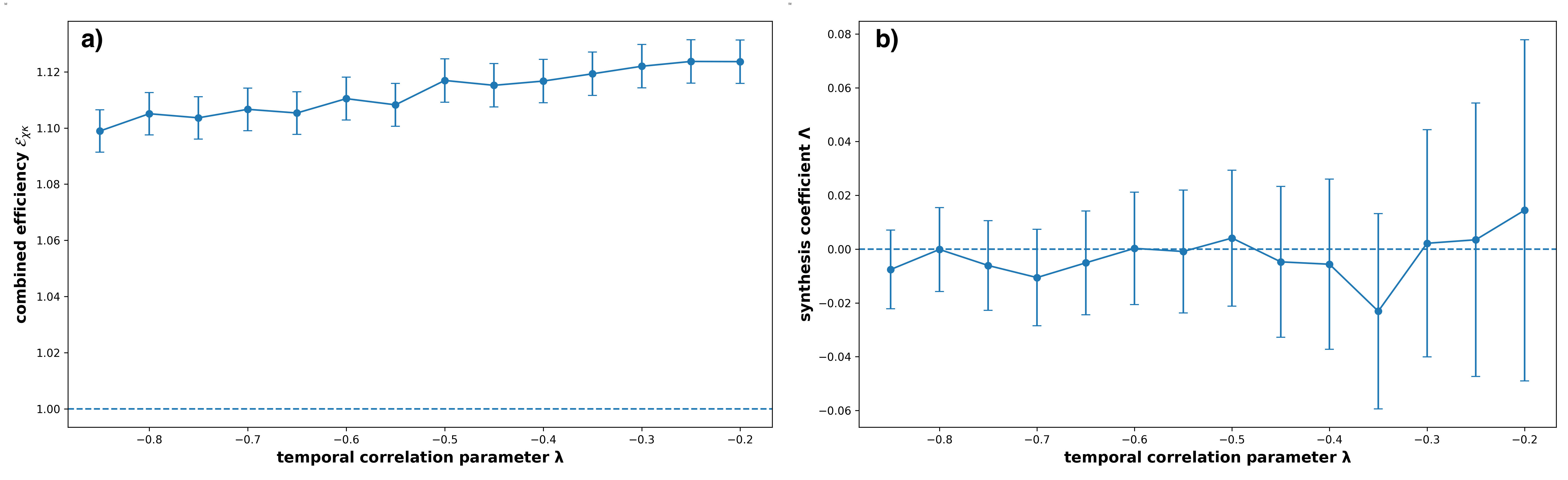}
\caption{
Efficiency gain and synthesis diagnostic for temporal anti-correlation combined with TOW conservation.
(a) Combined efficiency
\({\cal E}_{\chi\kappa}=R_{00}/R_{\chi\kappa}\)
as a function of the temporal correlation parameter \(\lambda\)
for the representative setting
\(P_A=0.7\), \(P_B=0.5\), \(\sigma=1.3\), \(\eta=0.5\), and \(\kappa=1\).
Values above unity indicate reduced regret relative to the uncorrelated one-sided baseline.
The high-statistics data show a robust efficiency gain over the tested range of negative temporal correlations.
Error bars show propagated standard errors obtained from the standard errors of the regret estimates.
(b) Synthesis coefficient \(\Lambda\), which isolates the non-additive interaction between temporal anti-correlation and conservation-induced option-wise anti-correlation.
Because \(\Lambda\) is computed from differences of efficiency ratios, its uncertainty is larger than that of \({\cal E}_{\chi\kappa}\).
The panel is therefore used as an interaction diagnostic rather than as a pointwise determination of a single optimal \(\lambda\).
}
\label{fig:synthesis}
\end{figure}

Figure~\ref{fig:synthesis} shows the corresponding efficiency gain and synthesis diagnostic with propagated standard errors.
The combined TOW system robustly improves efficiency relative to the uncorrelated baseline over the tested range of negative temporal correlations.
The improvement is not explained by temporal anti-correlation alone or by TOW conservation alone; rather, it reflects the conversion of nonredundant temporal sampling into differential evidence through the conservation-induced update.
The synthesis coefficient \(\Lambda\) is used as a diagnostic of this non-additive interaction, not as a pointwise determination of a sharply defined optimum in \(\lambda\).

The temporal-only effect discussed in Appendix C provides an independent baseline for anti-redundant exploration: even without TOW coupling, temporal anti-correlation can reduce redundant sampling.  For example, the tent-map-like estimate in Appendix C gives a search-time reduction to \(3T_0/4\) relative to an independent-sampling baseline.  By contrast, \(\Lambda\) measures the additional non-additive synthesis that appears only when this temporal ordering is combined with conservation-induced option-wise coupling.

\section{Scaling estimate for the optimal temporal correlation}

\begin{figure}[!t]
\centering
\includegraphics[width=0.98\textwidth]{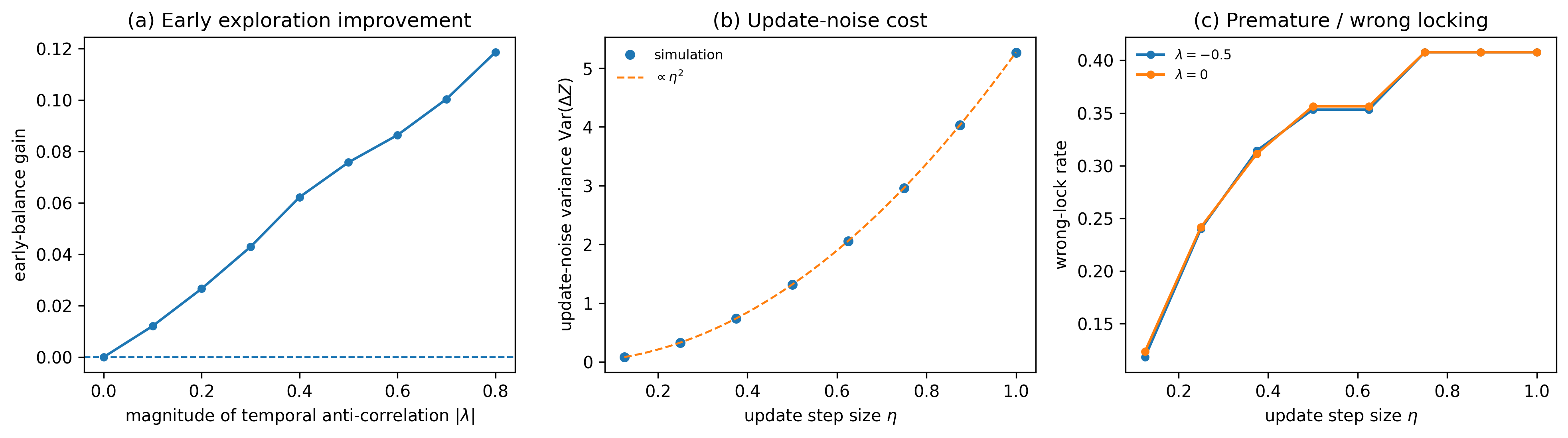}
\caption{
Internal mechanism behind the optimum.
(a) Temporal anti-correlation improves early exploration balance during the phase in which the drive can still affect the decision boundary.
(b) The variance of the differential update obeys the expected TOW scaling \(\mathrm{Var}(\Delta Z)\propto\eta^2\), confirming the update-noise cost in Eq.~\eqref{eq:update_noise}.
(c) Increasing \(\eta\) also increases premature or wrong locking, showing why the transduction from exploration to evidence formation has a finite window rather than growing monotonically with update strength.
}
\label{fig:mechanism}
\end{figure}

The intermediate optimum can be understood by balancing exploration gain against update-noise cost.  Temporal anti-correlation improves the early balance of samples between alternatives.  Let
\begin{equation}
B_{\rm early}=
1-\frac{|N_A-N_B|}{N_A+N_B}
\label{eq:early_balance}
\end{equation}
be the early sampling balance, evaluated during the phase \(|Z_t|<\sigma\) in which the exploratory signal can affect the decision boundary.  Internal diagnostics show that \(B_{\rm early}\) increases with \(|\lambda|\), and the corresponding smoothed estimation-variance proxy decreases.  This is the exploration-gain side of the mechanism.
The diagnostics in Fig.~\ref{fig:mechanism}(a) show this anti-redundancy gain directly.  The same figure also separates the two cost mechanisms: the TOW differential-update noise, Fig.~\ref{fig:mechanism}(b), and premature or wrong locking at large update size, Fig.~\ref{fig:mechanism}(c).

The cost side follows from the TOW update.  For a selected arm with reward probability \(P\), the update variable has variance
\begin{equation}
\mathrm{Var}(d_t)=\eta^2 P(1-P)(1+\omega_0)^2.
\end{equation}
Using Eq.~\eqref{eq:dZ}, the differential update noise scales as
\begin{equation}
\mathrm{Var}(\Delta Z)
\sim
\eta^2(1+\kappa)^2(1+\omega_0)^2.
\label{eq:update_noise}
\end{equation}
This scaling is directly confirmed by internal diagnostics: \(\mathrm{Var}(\Delta Z)\) changes approximately as \(\eta^2\) over the tested range.

We therefore introduce the dimensionless update-noise ratio
\begin{equation}
{\cal N}=
\frac{\eta(1+\kappa)(1+\omega_0)}{\Delta P}.
\label{eq:N}
\end{equation}
This quantity compares the characteristic scale of the TOW differential-update noise with the reward contrast that drives the correct decision.

A simple way to estimate the optimum is to regard temporal anti-correlation as producing an approximately linear anti-redundancy gain for small \(|\lambda|\), while the cost of using this temporal drive is controlled by the variance of the TOW differential update.  To lowest order, the usable range of temporal anti-correlation is therefore suppressed by the squared ratio between the update-noise scale and the reward contrast.  This gives the scaling form
\begin{equation}
|\lambda^\ast|
\simeq
\frac{\lambda_0}{1+c{\cal N}^2},
\label{eq:optimal_formula}
\end{equation}
where the constant \(c\) absorbs details of the early exploration window and of the decision threshold.


Equation~\eqref{eq:optimal_formula} should not be interpreted as an exact predictor of a unique optimal \(\lambda^\ast\).  Its role is to identify the leading control parameter that limits the constructive synthesis between temporal anti-redundancy and TOW-induced differential evidence formation.  The central point is therefore not an exact collapse of \(|\lambda^\ast|\) onto a single scalar variable, but the physical balance: temporal anti-correlation improves the nonredundancy of early exploration, whereas the TOW update amplifies both useful differential evidence and update noise.

The magnitude of the interaction also depends on a transduction window.  If \(\eta\) is too small, evidence formation is slow and temporal exploration is not efficiently converted into \(Z_t\).  If \(\eta\) is too large, the same temporal ordering can be amplified into premature evidence accumulation, causing wrong locking as shown in Fig.~\ref{fig:mechanism}(c).  Thus even in a stationary environment there is a trade-off between anti-redundant exploration and excessive commitment to noisy early observations.
We therefore write the qualitative structure as
\begin{equation}
\Lambda \sim
W_{\rm trans}(\eta,\sigma)
\frac{1}{1+c{\cal N}^2},
\label{eq:window}
\end{equation}
where \(W_{\rm trans}\) is not specified as a fitted universal function.
It represents the effective window in which temporal exploration is converted into differential evidence by the TOW update.
The proposed scaling is therefore expected to describe the leading dependence on update-noise scale and reward contrast, while the detailed optimum can also depend on the decision threshold, horizon, and the transduction window.
A discretized finite-state formulation of the minimal model is given in Appendix B.

\section{Discussion}

The present result should be understood as a minimal physical principle for efficient search, not as a claim about worst-case computational complexity.  We do not claim that NP-hard problems become easy in general, nor that the same scaling formula applies unchanged to SAT, TSP, or other high-dimensional search problems.  Rather, the result identifies a mechanism by which physical dynamics can reduce exploration cost: physical constraints and fluctuations generate correlations that are costly to reproduce by independent sampling.

The TOW dynamics provides a reduced setting in which this mechanism can be separated and measured.  Temporal correlations suppress redundant short-time exploration, while the TOW conservation law converts local reward feedback into an anti-correlated update of competing alternatives.  In this sense, a physical search system can be viewed as a correlated transport process on an effective search space: temporal structure reduces revisits, whereas physical constraints project local information onto collective update directions.  Efficient search emerges when the time scale of exploratory fluctuations is matched to the spatial scale over which constraints redistribute information.

This interpretation also provides a useful lens for more complex physical search systems.  Amoeba-inspired TSP and SAT dynamics contain time-dependent fluctuations together with spatial redistribution or constraint propagation, but their detailed rules are far more complex than the two-armed TOW dynamics and are not analyzed here.  The present analysis suggests that what may be generic is not the particular TOW scaling formula, but the structure of the mechanism: temporally structured exploration must be compatible with the constraint-induced redistribution that converts local information or local conflict into coordinated motion across degrees of freedom.

The two-armed setting should be understood as the minimal sector of broader multi-armed TOW dynamics.  A distinctive property of TOW dynamics is that, even in a \(K\)-armed bandit problem, the optimal penalty parameter is determined by the reward probabilities of the best and the second-best arms.  In this sense, the relevant competition is effectively reduced to the leading pair, with
\begin{equation}
  \gamma_{\rm eff}=P_{(1)}+P_{(2)},
\end{equation}
\begin{equation} 
\omega_0^{\rm eff}=\frac{\gamma_{\rm eff}}{2-\gamma_{\rm eff}},
\end{equation}
where \(P_{(1)}\) and \(P_{(2)}\) denote the largest and second-largest reward probabilities \cite{TOWFluid2016}.  This leading-pair reduction is a characteristic feature of TOW dynamics: lower-ranked arms influence transient exploration, but the asymptotic discrimination scale is governed by the gap and update balance between the two leading alternatives.

For high-dimensional physical search systems such as graph-based TSP or SAT-like dynamics, additional spatial effects may enter through network topology, coupling range, and relaxation time of the physical substrate.  Thus the present scaling variable should be viewed as the leading-pair core of a broader multi-armed extension.  A full high-dimensional theory would require matching the temporal exploration scale not only to the local TOW update scale, but also to the spatial propagation or decay scale of constraint-induced correlations.

Non-stationary environments would introduce an additional drift or switching time scale.  The stationary analysis developed here is expected to apply when the environmental change time is long compared with the evidence-formation time of the TOW dynamics.  When these time scales become comparable, strong temporal anti-correlation or large update steps may hinder adaptation by increasing premature locking to outdated evidence.  This clarifies the expected failure mode outside the stationary regime: if the best arm changes before the accumulated differential evidence can relax, the same mechanism that accelerates evidence formation in a stationary environment can amplify obsolete evidence.  In such cases the useful correlation time should be shortened, or the update scale reduced, so that the evidence-formation time remains below the environmental switching time.  The principle then becomes a three-way matching problem among exploratory correlation time, TOW update time, and environmental change time.

The broader contribution is therefore not a closed-form law for the exact optimum of every physical search system.  It is a decomposition of efficient physical search into two mechanisms: temporal anti-redundancy, which suppresses redundant sampling, and constraint-induced differential evidence formation, which converts local feedback into coordinated updates.  Efficient search emerges when these two mechanisms are matched.
This provides a design principle for physical search devices: fluctuations should not be maximized, but tuned to the scale at which physical constraints convert local observations into collective evidence.

\section{Conclusion}

We introduced a correlation-matching principle for TOW-based physical search.  In the minimal TOW bandit model, temporal anti-correlation and conservation-induced option-wise anti-correlation can be separated, varied, and measured.  Their constructive interaction occurs at intermediate temporal anti-correlation, not at maximum anti-correlation.

A scaling estimate identifies the dimensionless update-noise ratio
\begin{equation}
{\cal N}=\eta(1+\kappa)(1+\omega_0)/\Delta P
\end{equation}
as the leading parameter that limits how strongly temporal anti-correlation can be used.
This estimate is not intended as an exact predictor of a unique optimum; rather, it explains why stronger temporal anti-correlation is not necessarily better when temporal anti-redundancy is converted into TOW-induced differential evidence.

The central result is therefore a mechanistic decomposition of efficient physical search into temporal anti-redundancy and constraint-induced differential evidence formation.
Internal diagnostics support this mechanism: temporal anti-correlation improves early sampling balance, whereas the TOW update noise scales as \(\eta^2\).
The broader lesson is that physical search acceleration is not a consequence of randomness alone.
It arises when temporal exploration is matched to the spatial update scale imposed by conservation or constraint-induced coupling.

\section*{Acknowledgments}

This work was supported by SOBIN Institute LLC under Research Grant SP012.
The author acknowledges helpful discussions with Prof. Takuma Akimoto (Tokyo University of Science) in the early stage of this study.
The author used ChatGPT (OpenAI) for English editing and takes full responsibility for the final version.

\appendix

\section{Parameter sweeps for the scaling estimate}
\label{app:scaling_sweeps}

\begin{figure}[!t]
\centering
\includegraphics[width=0.98\textwidth]{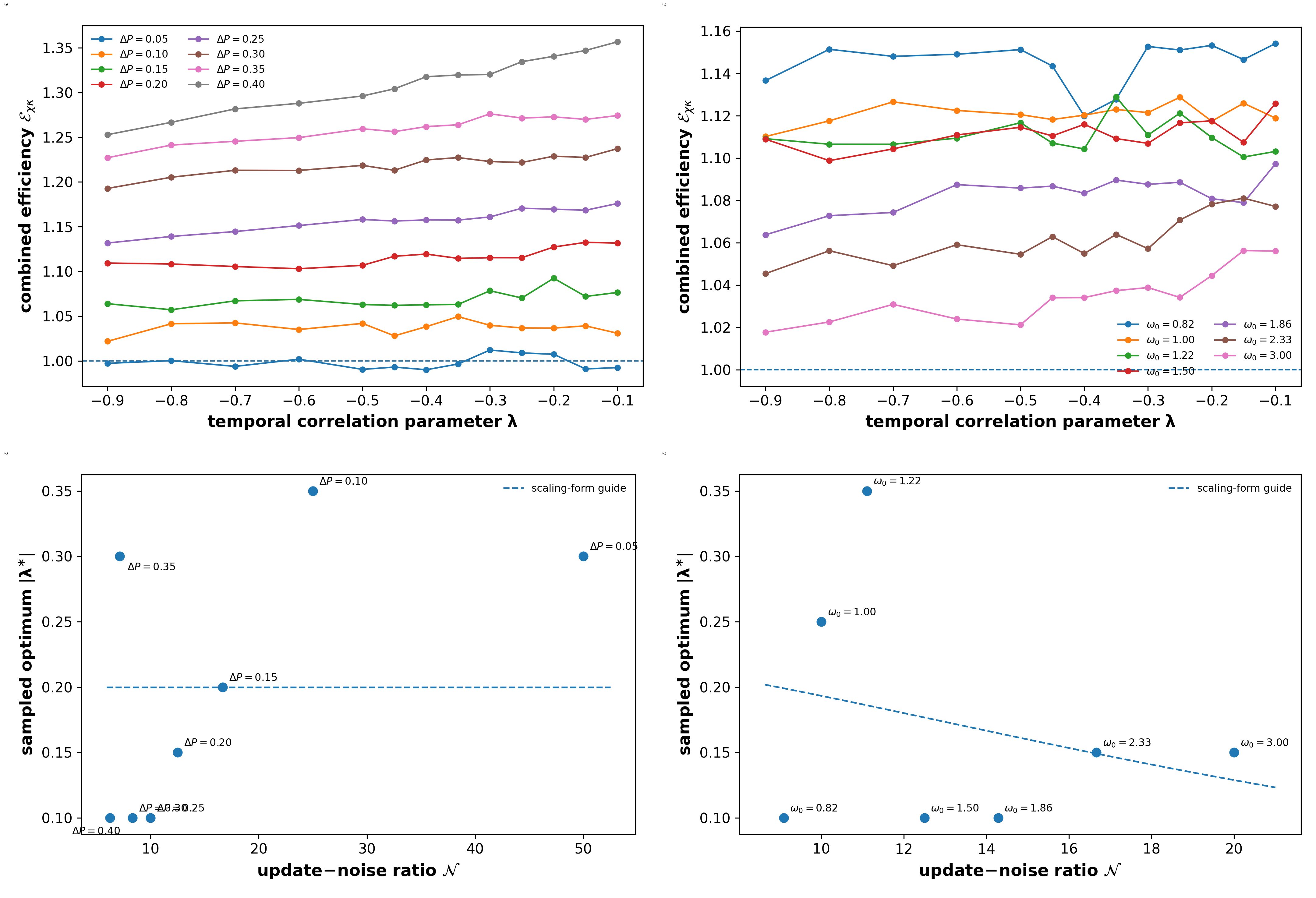}
\caption{
Parameter sweeps supporting the scaling interpretation.
(a) Changing the reward contrast \(\Delta P\) at fixed \(\omega_0=1.5\)
reshapes the efficiency landscape over temporal correlations.
(b) Changing the non-reward penalty \(\omega_0\) at fixed \(\Delta P=0.20\)
also modifies the efficiency landscape.
(c,d) Extracted optima \(|\lambda^\ast|\) plotted against the dimensionless update-noise ratio \({\cal N}=\eta(1+\kappa)(1+\omega_0)/\Delta P\). 
Here \({\cal N}\) is the update-noise-to-reward-contrast ratio: large \({\cal N}\) means that the physical update scale is large relative to the reward difference that must be resolved.
Because the optima are extracted from a finite \(\lambda\)-grid and the efficiency landscape can be broad near its maximum, the lower panels are used as scaling diagnostics rather than as a claim of an exact universal collapse.  Dashed curves are guides to the eye based on Eq.~\eqref{eq:optimal_formula}, not fitted universal laws.
}
\label{fig:scaling}
\end{figure}

Figure~\ref{fig:scaling} shows parameter sweeps used to diagnose the scaling estimate in Eq.~\eqref{eq:optimal_formula}.  These sweeps are not intended as a universal collapse plot.  Rather, they show how changing the reward contrast \(\Delta P\) and the non-reward penalty \(\omega_0\) reshapes the efficiency landscape over the temporal correlation parameter \(\lambda\).  The extracted optima are discrete because \(\lambda\) is sampled on a finite grid and because the efficiency landscape can be broad near its maximum.  They are therefore used as scaling diagnostics for the proposed update-noise-to-contrast ratio
\[
{\cal N}=\eta(1+\kappa)(1+\omega_0)/\Delta P,
\]
not as evidence for an exact collapse of \(|\lambda^\ast|\) onto \({\cal N}\).
In the lower panels, the dashed curves are guides to the eye based on the scaling form in Eq.~\eqref{eq:optimal_formula}; they are not used as evidence for a universal fit.
The small-\({\cal N}\) region should be interpreted with the same caution.
When the update-noise cost is weak, the usable anti-correlation is no longer limited primarily by \({\cal N}\), but by the finite early-exploration window, the decision boundary, and the sampled \(\lambda\)-grid.
The dashed curves therefore serve only as scaling-form guides for the onset of update-noise limitation, not as quantitative fits over the entire range.

\section{Discretized finite-state formulation}
\label{app:finite_state}

For completeness, we describe a discretized finite-state formulation of the minimal model.  Let the differential coordinate be represented on a finite grid
\begin{equation}
Z_t\in\{-N\delta_Z,(-N+1)\delta_Z,\ldots,N\delta_Z\},
\label{eq:app_finite_Z}
\end{equation}
where \(\delta_Z>0\) is the grid spacing, and let \(s_t\in\{-1,+1\}\).  The state is \(v_t=(Z_t,s_t)\).  The action is determined by Eq.~\eqref{eq:decision_rule}, with random choice at the zero boundary.  The signal transition is given by Eq.~\eqref{eq:signal}.  Given the action, rewards are drawn from
\begin{equation}
\Pr(R_t=1|a_t=A)=P_A,
\qquad
\Pr(R_t=1|a_t=B)=P_B.
\label{eq:app_reward_prob}
\end{equation}
The update can be written as
\begin{equation}
Z_{t+1}
=
\Pi_N\left[
\rho Z_t+
\eta(1+\kappa)u(a_t,R_t)
\right],
\label{eq:app_finite_update}
\end{equation}
where \(\Pi_N\) denotes projection onto the finite grid in Eq.~\eqref{eq:app_finite_Z}, including projection to the nearest grid point when necessary.  The signed update variable is
\begin{equation}
u(a_t,R_t)=
\begin{cases}
+1, & a_t=A,\ R_t=1,\\
-\omega_0, & a_t=A,\ R_t=0,\\
-1, & a_t=B,\ R_t=1,\\
+\omega_0, & a_t=B,\ R_t=0.
\end{cases}
\label{eq:app_u}
\end{equation}
Let \(\pi_t(v)=\Pr(v_t=v)\) be the distribution over finite states before trial \(t+1\).  The discretized dynamics defines a transition matrix \(M\) such that
\begin{equation}
\pi_{t+1}=M\pi_t .
\label{eq:app_transition_matrix}
\end{equation}
If \(P_A>P_B\), the regret up to horizon \(H\) is
\begin{equation}
R(H)
=
(P_A-P_B)\sum_{t=0}^{H-1}\Pr(a_t=B),
\label{eq:app_regret_prob}
\end{equation}
where \(a_t\) denotes the action selected from the state distribution \(\pi_t\).  Defining \(c(v)=1\) for states that select \(B\) and \(c(v)=0\) otherwise, with \(c(v)=1/2\) at the zero boundary if the boundary choice is randomized, this becomes
\begin{equation}
R(H)
=
(P_A-P_B)\sum_{t=0}^{H-1}c^\top M^t\pi_0 .
\label{eq:app_regret_matrix}
\end{equation}
Thus the model can be represented exactly as a finite-state Markov process after discretization.  The scaling theory in the main text should be viewed as a physical approximation to this discretized finite-state dynamics, not as a replacement for it.

\section{Temporal anti-correlation as anti-redundant search}
\label{app:temporal_antiredundancy}

The following argument should be read as a minimal first-passage-like estimate rather than an exact search-time theorem \cite{Redner2001}.
The main text focuses on the interaction between temporal exploration and the TOW conservation-induced update.  Here we isolate the temporal component alone and explain why temporal anti-correlation can improve search efficiency even before the TOW mechanism is introduced.

Consider first independent random sampling on a symmetric one-dimensional search domain.  At each trial, the searcher samples either side of the domain without memory of the previous trial.  Such a process explores, but it also has a simple inefficiency: it can repeatedly sample the same side while leaving the opposite side untested.  

A transparent artificial example is a mirror sequence.  Let
\begin{equation}
X_n \sim U(0,K),
\end{equation}
and define the search sequence as
\begin{equation}
X_n,\quad -X_n .
\end{equation}
This is not a chaotic sequence.  It is an idealized limiting case used only to display the mechanism.  Once one side of the domain is sampled, the opposite side is sampled immediately afterward.  Therefore the two consecutive samples are maximally anti-correlated with respect to the left--right search variable.  The marginal distribution of sampled positions is not broadened; the gain comes from suppressing immediate redundant sampling.

To express this idea generally, let \(C_s(\tau)\) be the autocorrelation of the task-relevant coarse-grained search variable.  We define the temporal anti-redundancy measure
\begin{equation}
\chi
=
-\sum_{\tau\ge1}w_\tau C_s(\tau),
\qquad
w_\tau\ge0,
\label{eq:app_chi}
\end{equation}
where the weights \(w_\tau\) specify which lags matter for the search task.  Positive \(\chi\) means that the temporal structure tends to suppress repeated sampling of recently explored alternatives or regions.

As a simple effective estimate, suppose that independent sampling has useful hit probability \(p_0\) per trial.  If temporal anti-correlation increases the rate of nonredundant trials by a factor \(1+\chi\), then
\begin{equation}
p_{\rm eff}\simeq p_0(1+\chi),
\label{eq:app_peff}
\end{equation}
and the corresponding mean search time scales as
\begin{equation}
T_{\rm eff}
\simeq
\frac{T_0}{1+\chi}, 
\label{eq:app_Teff}
\end{equation}
where \(T_0\) denotes the mean search time of the independent-sampling baseline.
This is not a universal exact law.  It is a minimal anti-redundancy estimate: temporal anti-correlation improves search when it converts trials that would have been redundant into trials that cover unexplored alternatives.

A tent-map-like symbolic observable gives a concrete nontrivial example.
Suppose the relevant binary search observable has autocorrelation
\begin{equation}
C_s(\tau)=\frac{(-1)^\tau}{2^\tau}.
\label{eq:app_tent_corr}
\end{equation}
Then
\begin{equation}
\sum_{\tau=1}^{\infty}C_s(\tau)
=
-\frac{1}{3},
\qquad
\chi_{\rm tent}
=
-\sum_{\tau=1}^{\infty}C_s(\tau)
=
\frac{1}{3}.
\label{eq:app_tent_chi}
\end{equation}
Substituting this value into Eq.~\eqref{eq:app_Teff} gives
\begin{equation}
T_{\rm tent}
\simeq
\frac{T_0}{1+1/3}
=
\frac{3}{4}T_0.
\label{eq:app_tent_gain}
\end{equation}
Thus, in this minimal estimate, the tent-map-like temporal structure reduces the search time by suppressing redundant sampling.

The numerical factor \(3/4\) is not universal; it depends on the observable, geometry, and weighting \(w_\tau\).  The point is that deterministic temporal structure can act as an exploration resource when it creates task-relevant anti-correlation.  This temporal-only mechanism is distinct from the TOW mechanism in the main text, where temporal anti-correlation interacts with conservation-induced option-wise anti-correlation.

The temporal-only estimate changes the rate at which nonredundant trials are generated, whereas the main text concerns an interaction term \(I(\chi,\kappa)\) that is absent when \(\kappa=0\).  Thus Appendix C explains one component of the mechanism; it does not account for the constructive synthesis measured by \(\Lambda\).

\bibliographystyle{unsrtnat}
\bibliography{references}

\end{document}